\documentclass{moriond}

\def\be{\begin{equation}}
\def\ee{\end{equation}}
\def\bea{\begin{eqnarray}}
\def\eea{\end{eqnarray}}

\newcommand{\Photo}{}

\begin{document}
\vspace*{4cm}
\title{AGN constraints on neutrino-dark matter scattering}

\author{James M.\ Cline}

\address{CERN, Theoretical Physics Department, Geneva, Switzerland,
and 
Department of Physics, McGill University, 3600 rue
University, Montr\'eal, Qu\'ebec, Canada H3A2T8
}

\maketitle\abstracts{The IceCube collaboration has identified
neutrinos of energy $\sim 10-100$\,TeV from the blazar TXS 0506+056
and the active galaxy NGC 1068, which must have traveled through a
dense dark matter spike surrounding the supermassive black holes that
power the galactic nuclei.  We use this to set new constraints on 
dark matter-neutrino scattering, and interpret the results in terms of
a dark photon that couples to baryon minus lepton number.
}

{\bf 1. Introduction.}
In recent years, the first two known sources of extragalactic
neutrinos were identified by the IceCube collaboration: a single
event of energy 290\,TeV from the blazar TXS 0506+056 in 2018
\cite{IceCube:2018dnn}, and $\sim 80$ events of energies up to 
15\,TeV from the active galaxy nucleus (AGN) of NGC 1068 in 
2022 \cite{IceCube:2022der}.  In fact, both sources are AGNs, the
blazar being one whose jet happens to point toward Earth.  In both
cases, the identification was made by pointing the Cerenkov cones back
to the source, and correlating with electromagnetic observations that
showed the AGNs to be in a flaring phase at the time of emission.
In the case of NGC 1068, a neutrino spectrum going as $\sim
E_\nu^{-3}$ was observed.

\begin{figure}[t]
\centerline{\includegraphics[width=0.55\linewidth]{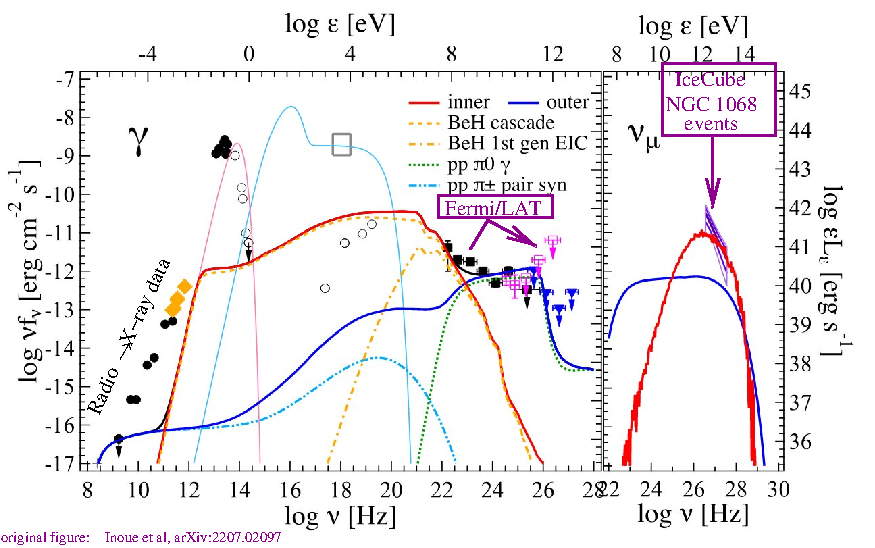}
\includegraphics[width=0.45\linewidth]{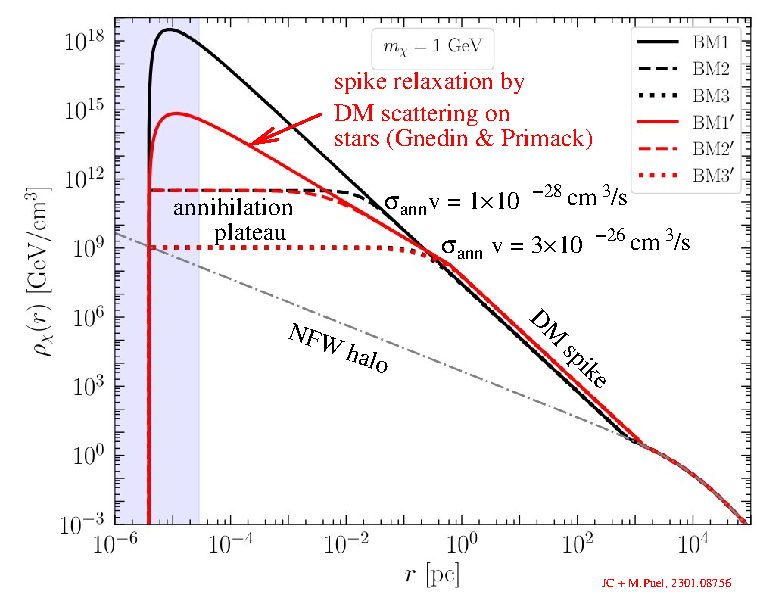}}
\caption[]{Left: example of predicted multimessenger signals from 
NGC 1068, with electromagnetic emission on the left and neutrino
flux on the right.  Adapted from Ref.\ \cite{Inoue:2022yak}.  Right:
examples of dark matter spike density profiles, illustrating the
possible softening of the inner region by DM annihilations or DM scattering
with stars.  Adapted from Ref.\ \cite{Cline:2023tkp}.  }
\label{fig:1}
\end{figure}

Numerous groups have developed models of the joint electromagnetic
(from radio to gamma-ray) and neutrino emissions from the relativistic
AGN jets; see for example
\cite{Murase:2011cy,Keivani:2018rnh,Cerruti:2021hah,Inoue:2022yak}.  
An example of the predicted spectra is shown in Fig.\ \ref{fig:1}
(left).  
Hadronic interactions such as 
$pN\to \pi^\pm X$, $N\gamma\to \pi^\pm X$ produce charged pions that
decay into neutrinos, while photons are produced by both hadronic and 
leptonic processes, including synchrotron, inverse Compton scattering,
pair production and annihilation \cite{Mastichiadis:2016gck}. 

\medskip
{\bf 2. Neutrino-dark matter interactions.}
If neutrinos scatter elastically with dark matter (DM), this could
attenuate the signal observed by IceCube.  This effect was previously
studied by Ref.\ \cite{Choi:2019ixb} to set limits on the $\nu$-DM
scattering cross section, considering only the DM in the halo and the
cosmological background.  However it is known that the supermassive
black hole that powers the blazar will have accreted a ``spike''
of DM, that is much more dense than the other components
\cite{Gondolo:1999ef}.  The precise form of the spike is
uncertain, depending on whether the DM could annihilate in the galactic
center, whether it experienced significant scattering with stars in this
region \cite{Gnedin:2003rj}, or if the galaxy underwent a recent
merger.  Despite these uncertainties, the spike is expected to reach
densities several orders of magnitude higher than the
Navarro-Frenk-White halo profile, which will lead to stronger limits
on the scattering cross section.

To quantify the attenuation of the neutrino signal, one can use a form
of the Boltzmann equation suitable for emission along the line of
sight, known as the cascade equation,\\
%\vskip-0.75cm
\centerline{\includegraphics[width=0.75\linewidth]{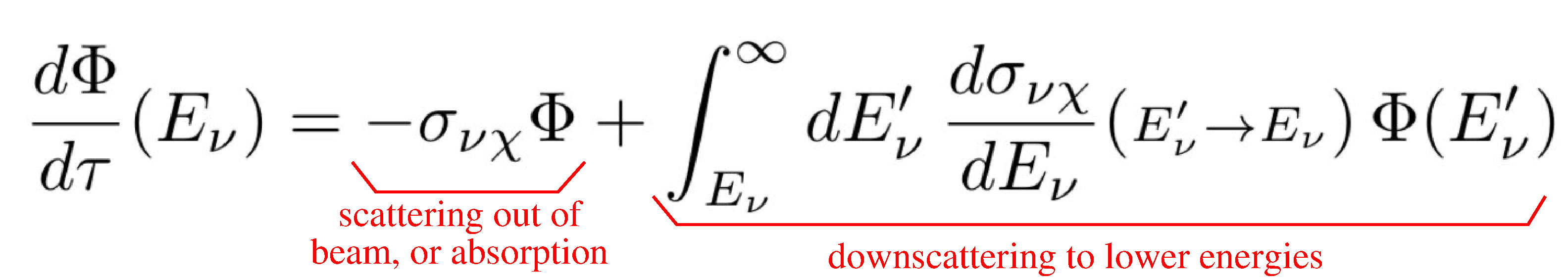}}
\noindent where $\tau$ is the accumulated column density per DM mass,
$\tau(r) = \int_{r_\nu}^r dr'\ \rho_\chi(r') / m_\chi$
and $r_\nu \sim 30\times R_S$ (the Schwarzschild radius) is where the neutrinos get
produced.

We applied this for the range of possible spikes shown in Fig.\
\ref{fig:1} to the TXS 0506+056 event in Ref.\ \cite{Cline:2023tkp}
(see also \cite{Ferrer:2022kei}), 
demanding at least $0.1$ IceCube event at $E_\nu =
290\,{\rm TeV}$ to derive the 90\%\,c.l.\ limits on $\sigma_{\nu\chi}$.
The resulting limits depend on how the cross section varies with
neutrino energy $E_\nu$.  We considered a simple power-law dependence
\be
	\sigma_{\nu\chi} = \sigma_0 \left(E/E_0\right)^n
\ee
with $n=0$ or $n=1$.  Low-energy scattering with a heavy mediator
of mass $m_{A'}$ and $E_\nu < m_\chi^2/m_{A'}$, where $m_\chi$ is the
DM mass, leads $n=1$, similarly to the standard model weak
interactions of neutrinos with nucleons.  For higher energies, the
cross section becomes constant, so these two choices have a natural 
physical origin. 

\begin{figure}[t]
\centerline{\includegraphics[width=0.5\linewidth]{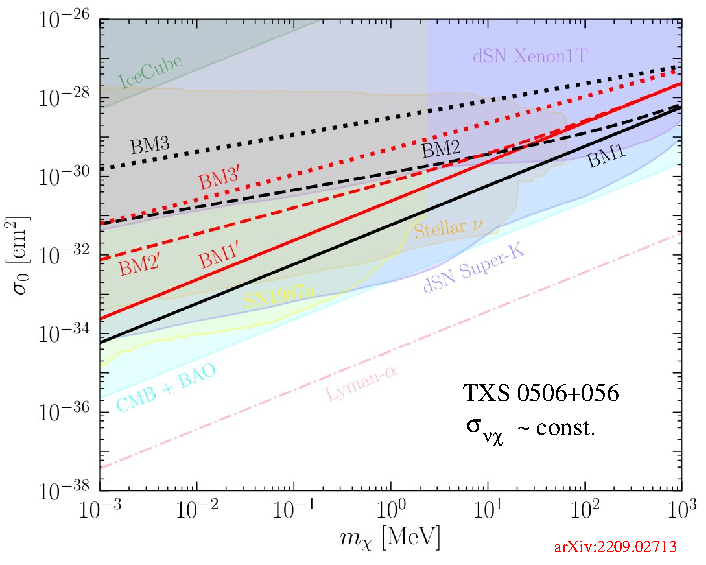}
\includegraphics[width=0.5\linewidth]{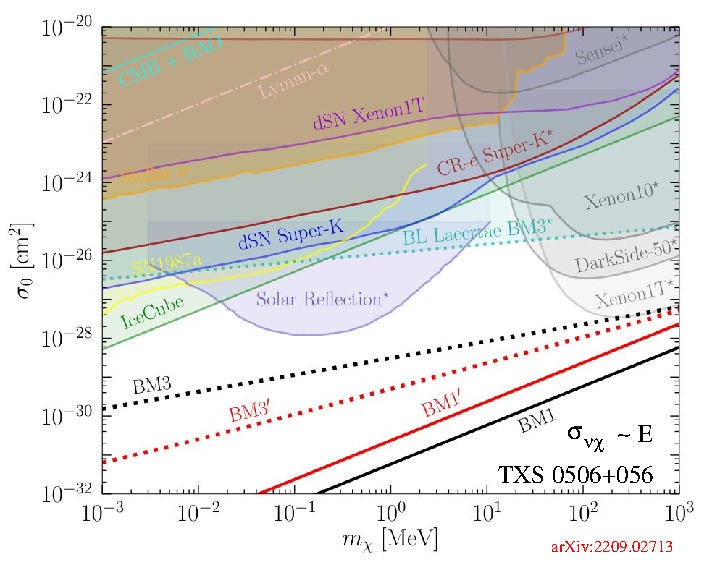}}
\caption[]{Upper limits on $\sigma_0$  versus DM mass $m_\chi$ for energy-independent (left)
and linear-in-energy (right) 
scattering from TXS 0506+056, derived in Ref.\ \cite{Cline:2022qld}, for the different
choices of DM spike densities shown in Fig.\ \ref{fig:1}. }
\label{fig:2}
\end{figure}

\medskip
{\bf 3. Model-independent results.}
 Our resulting limits\cite{Cline:2022qld} from TXS 0506+056 are shown
in Fig.\ \ref{fig:2} for the two possible energy dependences.  In the
background, previous constraints from various astrophysical and
cosmological sources are shown, which include effects on the CMB and
baryon acoustic oscillations, Lyman-$\alpha$ emissions, diffuse 
neutrinos from supernovae, detection of dark matter boosted by cosmic
neutrinos, and supernova 1987A.  The blazar constraints are not
competitive in this case.  For linear energy dependence, we must
rescale previous limits, which involve neutrino energies far below
$E_0=290\,$TeV, to the scale $E_0$ in order to compare.  In this case,
the blazar limit become world-leading, even for the most pessimistic
spike profiles.

\begin{figure}[t]
\centerline{\includegraphics[width=\linewidth]{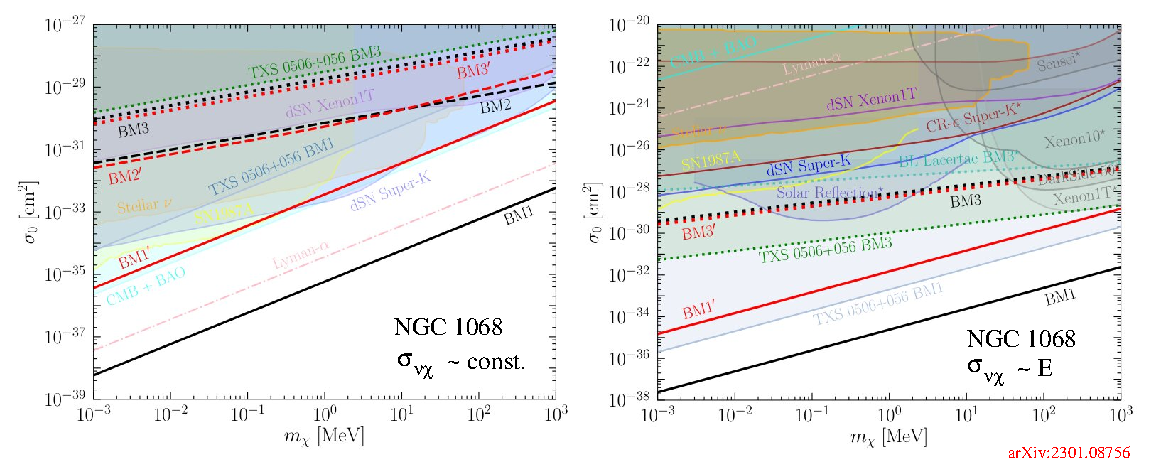}}
\caption[]{Upper limits on $\sigma_0$  versus DM mass $m_\chi$ for energy-independent (left)
and linear-in-energy (right) 
scattering from NGC 1068, derived in Ref.\ \cite{Cline:2023tkp}, for the different
choices of DM spike densities shown in Fig.\ \ref{fig:1}. }
\label{fig:3}
\end{figure}

To derive analogous limits from NGC 1068, we demanded at least 22
events should be observed in the energy window $E_\nu = [1,\,15]\,$TeV
to set the 90\%\,c.l.\,$\!$ limit.  Two null hypotheses were considered. First
we assumed that the standard model signal was the one observed by
IceCube, which leads to a limit independent of the neutrino emission
model.  Second we assumed that IceCube observed only a fraction of the
full emission due to scattering, taking phenomenological models such
as those cited previously
\cite{Murase:2011cy,Keivani:2018rnh,Cerruti:2021hah,Inoue:2022yak} 
to determine the predicted signal.  Both approaches give very similar
results (with the exception of one outlying model prediction).  The
resulting limits are shown in Fig.\ \ref{fig:3}, here using the
reference energy $E_0=10\,$TeV for rescaling the complementary
constraints when linear energy dependence is assumed.  Unlike TXS
0506+056, here the $n=0$ case can be more constraining, depending on
the spike profile, while the $n=1$ case can be less so, though still
generally stronger than other limits.  For $n=1$, there is interesting overlap
of the probed parameter space with values of $\sigma_0\sim
10^{-42.4}\,(m_\chi/{\rm MeV})^{4.5}$\,cm$^2$ and
$\sigma_0\sim 10^{-28.5}\,(m_\chi/{\rm MeV})$\,cm$^2$, respectively, that could explain the DM relic density through
freeze-in or freeze-out by $\chi\bar\chi\to \nu\bar\nu$
annihilation or its inverse.


\begin{figure}[t]
\centerline{\includegraphics[width=0.5\linewidth]{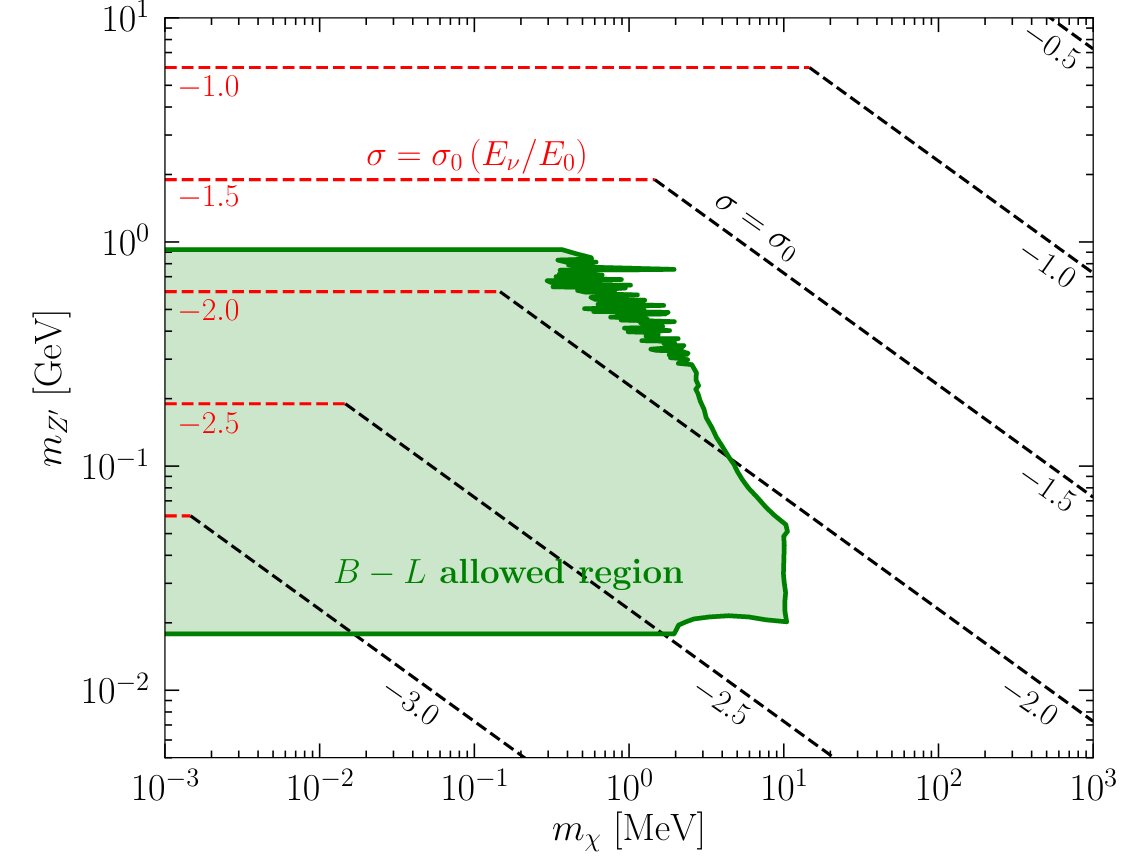}\includegraphics[width=0.5\linewidth]{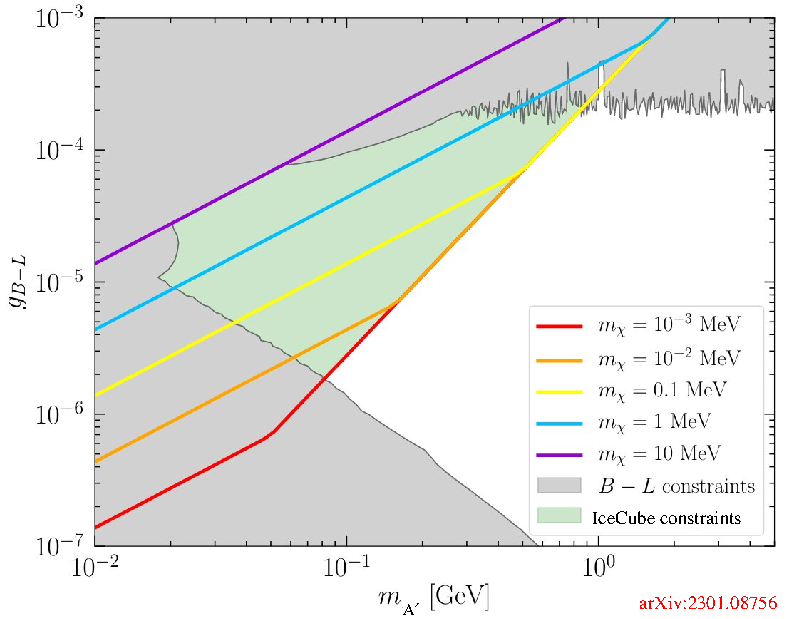}}
\caption[]{Left: upper limit contours of $\log_{10}g_{\rm eff}$ from
most stringent NGC 1068 constraint in
the $m_{A'}$-$m_\chi$ plane for the $B-L$ dark photon model; shaded
region is allowed by existing constraints.  Right: new NGC 1068 limits in the 
$g_\nu$-$m_{A'}$ plane corresponding to left panel, assuming
$g_\chi=1$.}
\label{fig:4}
\end{figure}

\medskip
{\bf 4. $\mathbf{B-L}$ dark photon interpretation.}
To make contact with a particle physics model of DM-$\nu$ scattering,
we supposed that the DM carries a $B-L$ (baryon minus lepton number)
charge, coupled to a massive dark photon $A'$, with coupling $g_\chi$
to DM and $g_\nu$ to neutrinos.  The cross section behaves as
$\sigma= g_{\rm eff}^4/(4\pi m_{A'}^2)$ for $E_\nu\gg m_{A'}^2/m_\chi$,
and $\sigma= g_{\rm eff}^4m_\chi E_\nu/(4\pi m_{A'}^4)$ for
$E_\nu\ll m_{A'}^2/m_\chi$, with $g_{\rm eff}^2 \equiv g_\chi g_\nu$.
Fig.\ \ref{fig:4} (left) shows the constraint contours of
$\log_{10}g_{\rm eff}$ (using the most restrictive DM spike profile), indicating the shaded region of $m_{A'}$-$m_{\chi}$
parameter space not already ruled out by existing limits.  Fig.\  
\ref{fig:4} (right) shows how these map into the $g_\nu$-$m_\chi$
plane, if $g_\chi=1$.  The limits depend on the choice of $m_\chi$,
and illustrate the potential to rule out new regions of the parameter space.

\medskip{\bf 5. Conclusions.}  The IceCube observations of neutrinos
from two AGNs give us competitive new limits on neutrino-DM
scattering.  Although they are uncertain because of astrophysics or
particle physics processes that can affect the DM spike profile near
the supermassive black hole, under reasonable assumptions they can be
stronger than other constraints.  This process can probe interesting
regions of realistic dark photon model parameter space, such as
coupling to baryon minus lepton number, and there is overlap with parameters that give the
right DM relic density.  Hopefully, this is just the beginning of 
neutrino AGN astronomy.

\medskip
{\bf Acknowledgments.}
This work was supported by the Natural Sciences and Engineering
Research Council (NSERC) of Canada.  I thank the CERN theory
department for its kind hospitality.

\section*{References}

\begin{thebibliography}{99}

%\cite{IceCube:2018dnn}
\bibitem{IceCube:2018dnn}
M.~G.~Aartsen \textit{et al.} [IceCube, Fermi-LAT, MAGIC, AGILE, ASAS-SN, HAWC, H.E.S.S., INTEGRAL, Kanata, Kiso, Kapteyn, Liverpool Telescope, Subaru, Swift NuSTAR, VERITAS and VLA/17B-403],
%``Multimessenger observations of a flaring blazar coincident with high-energy neutrino IceCube-170922A,''
Science \textbf{361}, no.6398, eaat1378 (2018)
doi:10.1126/science.aat1378
[arXiv:1807.08816 [astro-ph.HE]].
%864 citations counted in INSPIRE as of 10 Apr 2024

%\cite{IceCube:2022der}
\bibitem{IceCube:2022der}
R.~Abbasi \textit{et al.} [IceCube],
%``Evidence for neutrino emission from the nearby active galaxy NGC 1068,''
Science \textbf{378}, no.6619, 538-543 (2022)
doi:10.1126/science.abg3395
[arXiv:2211.09972 [astro-ph.HE]].
%235 citations counted in INSPIRE as of 10 Apr 2024

%\cite{Murase:2011cy}
\bibitem{Murase:2011cy}
K.~Murase, C.~D.~Dermer, H.~Takami and G.~Migliori,
%``Blazars as Ultra-High-Energy Cosmic-Ray Sources: Implications for TeV Gamma-Ray Observations,''
Astrophys. J. \textbf{749}, 63 (2012)
doi:10.1088/0004-637X/749/1/63
[arXiv:1107.5576 [astro-ph.HE]].
%221 citations counted in INSPIRE as of 29 Apr 2024


%\cite{Keivani:2018rnh}
\bibitem{Keivani:2018rnh}
A.~Keivani, K.~Murase, M.~Petropoulou, D.~B.~Fox, S.~B.~Cenko, S.~Chaty, A.~Coleiro, J.~J.~DeLaunay, S.~Dimitrakoudis and P.~A.~Evans, \textit{et al.}
%``A Multimessenger Picture of the Flaring Blazar TXS 0506+056: implications for High-Energy Neutrino Emission and Cosmic Ray Acceleration,''
Astrophys. J. \textbf{864}, no.1, 84 (2018)
doi:10.3847/1538-4357/aad59a
[arXiv:1807.04537 [astro-ph.HE]].
%241 citations counted in INSPIRE as of 29 Apr 2024

%\cite{Cerruti:2021hah}
\bibitem{Cerruti:2021hah}
M.~Cerruti, M.~Kreter, M.~Petropoulou, A.~Rudolph, F.~Oikonomou, M.~B\"ottcher, S.~Dimitrakoudis, A.~Dmytriiev, S.~Gao and S.~Inoue, \textit{et al.}
%``The Blazar Hadronic Code Comparison Project,''
PoS \textbf{ICRC2021}, 979 (2021)
doi:10.22323/1.395.0979
[arXiv:2107.06377 [astro-ph.HE]].
%9 citations counted in INSPIRE as of 29 Apr 2024

%\cite{Inoue:2022yak}
\bibitem{Inoue:2022yak}
S.~Inoue, M.~Cerruti, K.~Murase and R.~Y.~Liu,
%``Multimessenger emission from winds and tori in active galactic nuclei,''
PoS \textbf{ICRC2023}, 1161 (2023)
doi:10.22323/1.444.1161
[arXiv:2207.02097 [astro-ph.HE]].
%26 citations counted in INSPIRE as of 29 Apr 2024

%\cite{Cline:2023tkp}
\bibitem{Cline:2023tkp}
J.~M.~Cline and M.~Puel,
%``NGC 1068 constraints on neutrino-dark matter scattering,''
JCAP \textbf{06}, 004 (2023)
doi:10.1088/1475-7516/2023/06/004
[arXiv:2301.08756 [hep-ph]].
%14 citations counted in INSPIRE as of 29 Apr 2024





%\cite{Mastichiadis:2016gck}
\bibitem{Mastichiadis:2016gck}
A.~Mastichiadis,
%``Consequences of Proton Acceleration in Blazar Jets,''
Galaxies \textbf{4}, no.4, 59 (2016)
doi:10.3390/galaxies4040059
%0 citations counted in INSPIRE as of 10 Apr 2024


%\cite{Choi:2019ixb}
\bibitem{Choi:2019ixb}
K.~Y.~Choi, J.~Kim and C.~Rott,
%``Constraining dark matter-neutrino interactions with IceCube-170922A,''
Phys. Rev. D \textbf{99}, no.8, 083018 (2019)
doi:10.1103/PhysRevD.99.083018
[arXiv:1903.03302 [astro-ph.CO]].
%50 citations counted in INSPIRE as of 29 Apr 2024


%\cite{Gondolo:1999ef}
\bibitem{Gondolo:1999ef}
P.~Gondolo and J.~Silk,
%``Dark matter annihilation at the galactic center,''
Phys. Rev. Lett. \textbf{83}, 1719-1722 (1999)
doi:10.1103/PhysRevLett.83.1719
[arXiv:astro-ph/9906391 [astro-ph]].
%640 citations counted in INSPIRE as of 29 Apr 2024

%\cite{Gnedin:2003rj}
\bibitem{Gnedin:2003rj}
O.~Y.~Gnedin and J.~R.~Primack,
%``Dark Matter Profile in the Galactic Center,''
Phys. Rev. Lett. \textbf{93}, 061302 (2004)
doi:10.1103/PhysRevLett.93.061302
[arXiv:astro-ph/0308385 [astro-ph]].
%155 citations counted in INSPIRE as of 29 Apr 2024

%\cite{Cline:2022qld}
\bibitem{Cline:2022qld}
J.~M.~Cline, S.~Gao, F.~Guo, Z.~Lin, S.~Liu, M.~Puel, P.~Todd and T.~Xiao,
%``Blazar Constraints on Neutrino-Dark Matter Scattering,''
Phys. Rev. Lett. \textbf{130}, no.9, 091402 (2023)
doi:10.1103/PhysRevLett.130.091402
[arXiv:2209.02713].
%19 citations counted in INSPIRE as of 29 Apr 2024

%\cite{Ferrer:2022kei}
\bibitem{Ferrer:2022kei}
F.~Ferrer, G.~Herrera and A.~Ibarra,
%``New constraints on the dark matter-neutrino and dark matter-photon scattering cross sections from TXS 0506+056,''
JCAP \textbf{05}, 057 (2023)
doi:10.1088/1475-7516/2023/05/057
[arXiv:2209.06339 [hep-ph]].
%20 citations counted in INSPIRE as of 29 Apr 2024


\end{thebibliography}
\end{document}